\documentclass[12pt]{article}
\usepackage{amssymb}

\usepackage{graphicx}
\usepackage{amsmath}

\input{tcilatex}

\begin{document}

\title{\textbf{Heavy particle electroweak loop effects in extra-dimensional models
with bulk neutrinos}}
\author{{\Large T. P. Cheng}$^{\ast }$ {\Large and} {\Large Ling-Fong Li}$^{\dagger
} ${\Large \ } \\
$^{\ast }${\small Department of Physics and Astronomy, University of
Missouri, St. Louis, MO 63121}\\
$^{\dagger }${\small Department of Physics, Carnegie Mellon University,
Pittsburgh, PA 15213}}
\date{}
\maketitle

\begin{abstract}
One way to detect the presence of new particles in theories beyond the
standard model is through their contribution to electroweak loop effects. We
comment on the importance of a consistent inclusion of their mixing angles
to ensure that the physical requirement of heavy particle decoupling is
fulfilled. We illustrate our points by a detailed discussion of the lepton
flavor changing effect $\mu \rightarrow e\gamma ,$ investigated recently by
Kitano, in the Randall-Sundrum model. Our remarks are equally applicable to
models with large compactified dimensions where bulk neutrinos are
introduced to account for the observed neutrino oscillations.
\end{abstract}

\section{Introduction}

\noindent There is considerable interest in the attempts of solving the
gauge hierarchy problem in the context of theories invoking extra spatial
dimensions. One category of models assumes that the extra dimensions are
large. Mass hierarchy results from the large volume effect\cite{dimo}.
Another class makes use of a non-factorizable metric with a warp factor,
leading to exponential suppressions of Planck scale masses for the relevant
fields which are assumed to reside on the ``visible'' 3-brane\cite{RS}. Only
gravitons and, in some models, also other fields, can propagate in the extra
dimensional space. Such a bulk field will have a tower of Kaluza-Klein
states with ever increasing masses. These states often provide us with
definitive signatures of these extra dimension theories. If such bulk fields
can mix with ordinary SM fields, their presence can in principle be detected
through their contributions to the electroweak loop effects. There exist
already a substantial body of literature discussing the constraints on such
KK states by using the existing, or future, electroweak precision data\cite
{misc}. In this paper we wish to emphasize the importance of a proper and
complete inclusion of the mixing angle effects so that the physically
sensible requirement of the decoupling theorem can be satisfied.

We believe that the study of any physical phenomena at a given distance
scale should not depend sensitively on our knowledge of the physics on much
shorter scales, heavy particles should decouple from low-energy processes.
Namely, the effects of heavy particles in the virtual intermediate states
are suppressed by inverse powers of the heavy particle masses\cite
{app-carr-thm}. This comes about because the relevant amplitudes are reduced
by the heavy particle propagators. However, if the heavy mass comes from
spontaneous symmetry breaking, the corresponding Yukawa coupling is also
large and this can neutralize the large mass of the denominator. This may
lead to a violation of the decoupling theorem in the low energy effective
theory\cite{CL91}. For example, in the Standard Model, the $\rho $ parameter
grows with $m_{t}^{2}$ and is quite sensitive to the value of $m_{t}.$ In
fact, this is one of the clue to the $t$-quark mass before its discovery. On
the other hand, if the large mass can be attributed to a gauge invariant
mass term, then decoupling should be effective because here one does not
need a large Yukawa coupling. Often in a model, particles have a mixture of
bare and Yukawa-coupling-induced masses. These decoupling effects may show
up as mass-suppressed mixing angles. Masses and mixing angles are often
related because they all follow from the same (nondiagonal) mass matrices.

A proto-typical case is the seesaw mechanism for generating neutrino masses 
\cite{seesaw}: besides the ordinary light neutrinos with masses $m_{\nu
}\simeq \hat{m}^{2}/\hat{M},$ there is also at least one other superheavy
neutrino with a mass $m_{N}\simeq \hat{M}.\;$[The small mass is assumed to
have a magnitude comparable to the masses of ordinary quarks and charged
leptons, $\hat{m}\simeq 1\,GeV$, the large mass being an intermediate mass
scale below the Planck scale, $\hat{M}\simeq 10^{12}GeV$.] In the charged
weak current, the charged lepton is coupled to the combination $\left( \cos
\theta \left| \nu \right\rangle +\sin \theta \left| N\right\rangle \right) ,$
where $\left| \nu \right\rangle $ stands for some superposition of the light
neutrino states with mixing angles that are not necessarily small, but the
angle $\theta $ is mass-suppressed 
\begin{equation}
\theta \simeq \frac{\hat{m}}{\hat{M}}.  \label{seesaw-angle}
\end{equation}
Such a mixing angle simply reflects the property of the mass matrix that, in
the $\hat{M}\rightarrow \infty $ limit, the mixing of the singlet neutrino
goes to zero. The presence of non-zero neutrino masses naturally leads to
flavor violation loop effects such as $\mu \rightarrow e\gamma .\;$Both
light and heavy neutrinos contribute, leading to a branching ratio\cite
{CL-ps00} 
\begin{equation}
B\left( \mu e\gamma \right) =\frac{3\alpha }{8\pi }\zeta ^{2}\theta ^{4}
\label{seesaw-muegamma}
\end{equation}
where $\alpha $ is the fine structure constant. The factor $\zeta $ being
some product of the mixing coefficients among light neutrinos is not
expected to be particularly small. Had one not taken into account of the
fact that the heavy-light mixing angle $\theta $ is mass-suppressed, one
would erroneously conclude that the heavy neutrino did not decouple (and
thus giving rise to an unacceptably large branching ratio). But in this
representation of the neutrino states, decoupling manifest itself in the
form $\theta ^{4}=\left( \hat{m}^{2}/\hat{M}^{2}\right) ^{2}=m_{\nu
}^{2}/m_{N}^{2}$ which yields an immeasurably small branching ratio ---
because of the superheavy neutrino mass in the denominator, as well as the
tiny light neutrino mass in the numerator. The seesaw model of neutrino mass
is considered to be an attractive possibility because the presence of such
self-consistent features. We suggest that any physically sensible theory
containing superheavy particles would have this type of properties that
automatically ensures that the heavy states are decoupled in low energy
processes.

The mass suppressed mixing angle follows from a special feature of the
seesaw neutrino mass matrix --- the absence of Majorana mass terms for the
left handed neutrino and a superheavy entry for the right handed neutrino
mass term. In essence, the reason that the decoupling holds in this case is
due to the fact the large mass can be realized by having large bare mass ($%
\nu _{R}$ being a SM singlet) without having large Yukawa coupling. That
mass matrices have the structure which gives rise to decoupling is rather
common in models involving heavy particle states. Thus the question of mass
suppressed mixing angles is very important in our consideration of heavy
particle contribution to low energy loop effects. In this paper we shall
illustrate our points in the Randall-Sundrum model\cite{RS} with bulk
neutrinos\cite{GN}. The investigation of lepton flavor violation loop
effects in this context have recently been carried out by Kitano\cite{kitano}%
. Here we complete his discussion, in particular with respect the
possibility of extracting a meaningful bound on the heavy neutrino mass.

\section{Mixing angles in the RS model with bulk neutrinos}

The Randal-Sundrum model presupposes a five dimensional spacetime. The extra
spatial dimension is taken to be a compactified $S^{1}/Z_{2}$ orbifold with
a coordinate $y=r_{c}\phi ,$ with $r_{c}$ being the radius of the compact
dimension and the angle $\phi $ having a range of $\left[ -\pi ,\pi \right] $
with opposite sides identified. There are two 3-branes fixed at $\phi =\pi $
(the ``visible'' brane containing the SM fields) and at $\phi =0$ (the
``hidden'' brane, also called the Planck brane). The resultant metric is
non-factorizable: 
\begin{equation}
ds^{2}=e^{-2kr_{c}\left| \phi \right| }\eta _{\mu \nu }dx^{\mu }dx^{\nu
}-r_{c}^{2}d\phi ^{2}
\end{equation}
where $k$, the bulk curvature, has the order of fundamental mass scale $\hat{%
M}_{5}$ which is comparable to the Planck mass. The exponential warp factor $%
e^{-2kr_{c}\left| \phi \right| }$ causes a rescaling of the fields, which
changes any mass parameter\ in the fundamental theory ($\simeq $ Planck
scale) to an effective mass on the visible brane as $M=e^{-kr_{c}\pi }\hat{M}%
_{5}$\ ($\simeq $ electroweak scale $M_{EW}$). Namely, for a choice of $%
kr_{c}\approx 12,$ we can have 
\begin{equation}
\epsilon =e^{-kr_{c}\pi }\approx 10^{-16}\;\;\;\;\;\;\;\;M=\epsilon \hat{M}%
_{5}\approx 10^{3}GeV.
\end{equation}

Because this mechanism does not allow any intermediate scale, between the
Planck and electroweak, to appear, the seesaw mechanism for generating a
naturally small neutrino mass is not applicable in the original RS model. In
this connection, Grossman and Neubert\cite{GN} introduced a bulk fermion
field\footnote{%
Cancellation of parity anomaly requires that there be even number of bulk
fermions. Since the presence of multiple bulk fermions should not introduce
qualitative changes in our result, we shall ignore such complication and
stick with one bulk fermion.}. They have shown that, for a reasonable range
of parameters, the zero mode of such a fermion has a very small wavefunction
at the physical brane and the Higgs generated mass can also be naturally
suppressed. In this way, neutrino masses that are many orders smaller than $%
M_{EW}$\ can be obtained.

The bulk fermion (with mass $M_{b}$) has the Kaluza-Klein decomposition of 
\begin{equation}
\Psi _{5}^{L,R}\left( x,\phi \right) =\frac{e^{2kr_{c}\left| \phi \right| }}{%
\sqrt{r_{c}}}\sum_{n}\hat{f}_{n}^{L,R}\left( \phi \right) \psi
_{n}^{L,R}\left( x\right)  \label{KK}
\end{equation}
where the superscripts $\left( L,R\right) $ signify the chirality states $%
\Psi _{5}^{L,R}=%
{\frac12}%
\left( 1\mp \gamma _{5}\right) \Psi _{5},$ and $\left\{ \hat{f}%
_{n}^{L,R}\left( \phi \right) \right\} $ are the appropriate sets of
complete orthonormal functions (in this case some combinations of Bessel
functions) normalized so that $\psi _{n}\left( x\right) $ has the canonical
scale in four dimensions, 
\begin{equation}
S_{\psi }=\int d^{4}x\left\{ \bar{\psi}_{n}\left( x\right) i\NEG%
{\partial}\psi _{n}\left( x\right) -M_{n}\bar{\psi}_{n}\left( x\right) \psi
_{n}\left( x\right) \right\} .  \label{dirac}
\end{equation}
The KK fields $\psi _{n\neq 0}^{L,R}\left( x\right) $ has electroweak scale
masses $M_{n}=\epsilon kx_{n}$ with $x_{n}$ (corresponding to zeros of some
combinations of the Bessel functions) being of order one. The presence of
such states brings hope for experimental searches, or equivalently, for
severe constraints by known phenomenology. Our focus in this paper is the
proper accounting, in such analyses, of the important effects due to the
mixing angles between these heavy states and the SM fields.

Grossman and Neubert\cite{GN} have shown that bulk fermion zero modes $%
\left( x_{0}=0\right) $ exist. If we impose the orbifold symmetry $\phi
\rightarrow -\phi ,$ then only one of the chiral zero modes survives. Let it
be $\psi _{0}^{R}\left( x\right) ,$ which has a suppressed wavefunction on
the visible brane 
\begin{equation}
\hat{f}_{0}^{R}\left( \phi =\pi \right) \simeq \sqrt{\epsilon kr_{c}}%
\epsilon ^{\nu -%
{\frac12}%
}=O\left( \epsilon ^{\nu }\right) \;\;\;\;\;\;\;\text{with }\nu =\frac{M_{b}%
}{k}>\frac{1}{2}.  \label{n0-wf}
\end{equation}
where $M_{b}$ is the bulk fermion mass parameter in the original $5$%
-dimensional Lagrangian. Similarly, orbifold symmetry requires the
wavefunctions for the left-handed KK excitations, when evaluated on the
visible brane, to vanish\ $\hat{f}_{n}^{L}\left( \phi =\pi \right) =0$ while
those for the right-handed fields\ have values 
\begin{equation}
\hat{f}_{n\neq 0}^{R}\left( \phi =\pi \right) \simeq \sqrt{\epsilon kr_{c}}%
=O\left( \epsilon ^{%
{\frac12}%
}\right) .  \label{n-wf}
\end{equation}
Thus, $\left( \hat{f}_{0}^{R}/\hat{f}_{n\neq 0}^{R}\right) _{\phi =\pi
}=O\left( \epsilon ^{\nu -%
{\frac12}%
}\right) $ is quite small since $\epsilon $ is tiny and $\nu >\frac{1}{2}.$

Relevant to our discussion of neutrino mass matrix, we shall only display
the Yukawa interaction between the SM left-handed lepton doublet $%
L^{L}=\left( l^{L},\,\nu _{l}^{L}\right) ,$ the right-handed bulk fermion $%
\Psi _{5}^{R}$, and the Higgs doublet $H=\left( h^{+},h^{0}\right) ,$ with
its conjugate being $\tilde{H}=i\sigma _{2}H^{\ast }.\;$Again for simplicity
we shall suppress the lepton generation indices $\left( e,\mu ,\tau \right) $
at this stage. 
\begin{equation}
S_{Y}=-\int d^{4}x\epsilon ^{4}\hat{Y}_{5}\left\{ \bar{L}_{5}^{L}\left(
x\right) \tilde{H}_{5}\left( x\right) \Psi _{5}^{R}\left( x,\pi \right)
+h.c.\right\}  \label{5dYukawa}
\end{equation}
where the factor $\epsilon ^{4}$ originates from the square root of the
metric determinant, and $\hat{Y}_{5}$, the fundamental Yukawa coupling, is
dimensionful, expected to be somewhat less than $\hat{M}_{5}^{-%
{\frac12}%
};$ and the fundamental fields $L_{5}^{L}\left( x\right) $ and $\tilde{H}%
_{5}\left( x\right) $ can be replaced by the effective fields (which have
the canonical normalizations in the four dimensional spacetime):\ $\epsilon
^{-3/2}L\left( x\right) $ and $\epsilon ^{-1}H\left( x\right) ,$
respectively. After substituting in the KK decomposition of Eq.(\ref{KK}),
the Yukawa interaction in Eq.(\ref{5dYukawa}) has now the form: 
\begin{equation}
S_{Y}=-\int d^{4}x\left\{ y_{0}\bar{L}^{L}\left( x\right) \tilde{H}\left(
x\right) \psi _{0}^{R}\left( x\right) +\sum_{n=1}y_{n}\bar{L}^{L}\left(
x\right) \tilde{H}\left( x\right) \psi _{n}^{R}\left( x\right) +h.c.\right\}
\label{yukawa}
\end{equation}
where $y_{n}=\hat{Y}_{5}\hat{f}_{n}^{R}\left( \phi =\pi \right) /\sqrt{%
\epsilon r_{c}}$. After using the estimates of Eqs.(\ref{n0-wf}) and (\ref
{n-wf}), and $\sqrt{k}\hat{Y}_{5}\lesssim 1$, the four dimensional Yukawa
couplings have the size of 
\begin{equation}
y_{0}\lesssim \epsilon ^{\nu -%
{\frac12}%
}\;\;\;\;\;\text{and\ \ \ \ \ }y_{n}\lesssim 1.
\end{equation}
Spontaneous symmetry breaking due to the Higgs mechanism results in a
non-zero vacuum expectation value for the neutral scalar field $\left\langle
h^{0}\right\rangle =v$ of the electroweak scale. Eq.(\ref{yukawa}) leads to
mass terms : 
\begin{equation}
m_{0}\bar{\nu}_{l}^{L}\psi _{0}^{R}+\sum_{n=1}m_{n}\bar{\nu}_{l}^{L}\psi
_{n}^{R}
\end{equation}
with the ``Yukawa masses'' 
\begin{equation}
m_{0}=y_{0}v\lesssim \epsilon ^{\nu -%
{\frac12}%
}v\ll M_{EW}\;\;\;\;\;\;\text{and\ \ \ \ \ \ }m_{n}=y_{n}v\lesssim M_{EW}.\;
\end{equation}
Combining with the Dirac mass terms of the KK states $\sum_{n=1}M_{n}\bar{%
\psi}^{L}\psi _{n}^{R}$, these mass terms can be written in a matrix form 
\begin{equation}
\left( 
\begin{array}{cccc}
\bar{\nu}^{^{\prime }L}, & \bar{\psi}_{1}^{L}, & \bar{\psi}_{2}^{L}, & ...
\end{array}
\right) \left( 
\begin{array}{cccc}
m_{0} & m_{1} & m_{2} & .\,. \\ 
0 & M_{1} & 0 & .\,. \\ 
0 & 0 & M_{2} & .\,. \\ 
: & : & : & 
\end{array}
\right) \left( 
\begin{array}{c}
\psi _{0}^{R} \\ 
\psi _{1}^{R} \\ 
\psi _{2}^{R} \\ 
:
\end{array}
\right) .
\end{equation}

For simplicity, let us concentrate on the simplest nontrivial case by
cutting off the $n>1$ excitations, thus a neutrino mass term of $\bar{\Psi}%
_{L}^{\prime }\mathbb{M\Psi }_{R}^{\prime },$ where 
\begin{equation}
\bar{\Psi}_{L}^{\prime }=\left( 
\begin{array}{cc}
\bar{\nu}_{l}^{L}, & \bar{\psi}_{1}^{L}
\end{array}
\right) ,\;\;\;\;\;\mathbb{M=}\left( 
\begin{array}{cc}
m_{0} & m_{1} \\ 
0 & M_{1}
\end{array}
\right) ,\;\;\;\;\;\mathbb{\Psi }_{R}^{\prime }=\left( 
\begin{array}{c}
\psi _{0}^{R} \\ 
\psi _{1}^{R}
\end{array}
\right)  \label{two-state}
\end{equation}
with $m_{0}\ll m_{1}\lesssim M_{1.}$ The mass matrix can be diagonalized in
terms of the mass eigenstates $\left( 
\begin{array}{cc}
\nu , & N
\end{array}
\right) $ by unitary transformation matrices $\mathbb{U}\left( \theta
_{L}\right) $ and $\mathbb{V}\left( \theta _{R}\right) $ acting on the left-
and right-handed fields, respectively: 
\begin{eqnarray}
\mathbb{U}\left( 
\begin{array}{c}
\nu \\ 
N
\end{array}
\right) _{L} &=&\left( 
\begin{array}{c}
\nu _{l}^{L} \\ 
\psi _{1}^{L}
\end{array}
\right) \simeq \left( 
\begin{array}{c}
\;\,\nu _{L}+\theta _{L}N_{L} \\ 
-\theta _{L}\nu _{L}+N_{L}
\end{array}
\right)  \notag \\
\mathbb{V}\left( 
\begin{array}{c}
\nu \\ 
N
\end{array}
\right) _{R} &=&\left( 
\begin{array}{c}
\psi _{0}^{R} \\ 
\psi _{1}^{R}
\end{array}
\right) \simeq \left( 
\begin{array}{c}
\;\,\nu _{R}+\theta _{R}N_{R} \\ 
-\theta _{R}\nu _{R}+N_{R}
\end{array}
\right)  \label{LRtwostate}
\end{eqnarray}
so\ that 
\begin{equation}
\mathbb{UMV}^{\dagger }=\mathbb{M}_{diag}=\left( 
\begin{array}{cc}
m_{\nu } & 0 \\ 
0 & m_{N}
\end{array}
\right)
\end{equation}
with\ $m_{\nu }\simeq m_{0}$ being very small and $m_{N}\simeq M_{1}$ very
large. The mixing angle for the left-handed field $\theta _{L}$ should be
fairly small, while $\theta _{R}$ for the right-handed fields is even more
suppressed: 
\begin{equation}
\theta _{L}\simeq \frac{m_{1}}{M_{1}}\lesssim 1\;\;\;\;\text{and}%
\;\;\;\;\theta _{R}\simeq \frac{m_{0}m_{1}}{M_{1}^{2}}\lesssim O\left(
\epsilon \right) .  \label{mix-angles}
\end{equation}

Next we will examine in some detail how such mixing angles will figure in
the constraint that the electroweak loop effects, such as $\mu \rightarrow
e\gamma ,$ will place on the new physics. Obviously for this purpose, we
must have at least two distinctive lepton flavors: $\nu _{l}=\nu _{e}$, $\nu
_{\mu }.$ Thus the $m_{0}$ factor in (\ref{two-state}) is now a $2\times 2$
non-diagonal mass matrix, whose elements are of same order magnitude as
before. The gauge and mass eigenstates in (\ref{two-state}) and (\ref
{LRtwostate}) must be expanded minimally to sets of three states: 
\begin{equation}
\mathbb{U}\left( 
\begin{array}{c}
\nu _{1} \\ 
\nu _{2} \\ 
\nu _{3}
\end{array}
\right) _{L}=\left( 
\begin{array}{c}
\nu _{e}^{L} \\ 
\nu _{\mu }^{L} \\ 
\psi _{1}^{L}
\end{array}
\right) .
\end{equation}
The mass eigenstates $\left\{ \nu _{i}\right\} $ correspond to two light
neutrinos with masses $m_{\nu 1}$ and $m_{\nu 2}$, on the order of zero-mode
Yukawa mass $m_{0},$ and one heavy neutrino with $m_{\nu 3}\simeq M_{1}.$
(We have changed the label for the heavy neutrino from $N$ to $\nu _{3}.$)
For simplicity, we shall assume that the unitary matrix $\mathbb{U}$ can be
parametrized by two mixing angles: one being the rotation angle $\omega $ in
the $\left( 1,2\right) $-plane, and the other being $\theta _{L},$ the $%
\left( 2,3\right) $ light-heavy rotation angle. 
\begin{equation}
\mathbb{U}_{li}=\left( 
\begin{array}{ccc}
\cos \omega & -\cos \theta _{L}\sin \omega & \;\sin \theta _{L}\sin \omega
\\ 
\sin \omega & \;\cos \theta _{L}\cos \omega & -\sin \theta _{L}\cos \omega
\\ 
0 & \sin \theta _{L} & \cos \theta _{L}
\end{array}
\right)  \label{mixings}
\end{equation}
We now proceed to the discussion of the $\mu e\gamma $ loop effects.

\section{$\protect\mu \rightarrow e\protect\gamma $ : heavy particle in the
gauge boson loop}

The decay amplitude for the $\mu \left( p\right) \rightarrow e\left(
p-q\right) +\gamma \left( q\right) $ can be written as 
\begin{equation}
T\left( \mu e\gamma \right) =\frac{ie}{16\pi }\varepsilon _{\lambda }^{\ast
}\left( q\right) \bar{u}_{e}\left( p-q\right) \sigma ^{\lambda \rho }q_{\rho
}\left[ A_{+}\left( 1+\gamma _{5}\right) +A_{-}\left( 1-\gamma _{5}\right) %
\right] u_{\mu }\left( p\right) .
\end{equation}
The branching ratio (with $m_{e}=0$) can then be expressed in terms of the
invariant amplitudes $A_{\pm }$ as 
\begin{equation}
B\left( \mu e\gamma \right) =\frac{6e^{2}M_{W}^{4}}{g^{4}m_{\mu }^{2}}\left(
\left| A_{+}\right| ^{2}+\left| A_{-}\right| ^{2}\right) ,
\end{equation}
where $g$ is the weak gauge coupling, $M_{W}$ the weak gauge boson mass.

First we discuss the invariant amplitudes $A_{\pm }^{W}$ coming from the
gauge boson loop contribution $\mu ^{-}\rightarrow \left( \nu _{i}W_{\gamma
}^{-}\right) \rightarrow e^{-}$ where the photon is emitted by the charged $%
W $ boson in the loop (as denoted by the subscript $\gamma $). The gauge
boson coupling to the charged lepton and massive neutrinos is 
\begin{equation}
\mathcal{L}\left( Wl\nu _{i}\right) =\frac{g}{\sqrt{2}}\mathbb{U}_{li}\bar{l}%
\gamma ^{\alpha }%
{\frac12}%
\left( 1-\gamma _{5}\right) \nu _{i}W_{\alpha }^{-}+h.c.,
\end{equation}
leading (after a detailed calculation\cite{CL-ps00}) to the amplitudes of $%
A_{-}^{W}=0$ and 
\begin{equation}
A_{+}^{W}=\frac{g^{2}m_{\mu }}{8\pi M_{W}^{2}}\sum_{i=1}^{3}\mathbb{U}_{\mu
i}^{\ast }\mathbb{U}_{ei}F\left( \frac{m_{i}^{2}}{M_{W}^{2}}\right) ,
\label{W-amplitude}
\end{equation}
where the function 
\begin{equation}
F\left( z\right) =\frac{1}{6\left( 1-z\right) ^{4}}\left(
10-43z+78z^{2}-49z^{3}-18z^{3}\ln z+4z^{4}\right)
\end{equation}
has limits of $F\left( 0\right) =5/3$ and $F\left( \infty \right) =2/3,$
respectively$.$ In our case we have two light neutrinos $\nu _{1,2}$ (thus $%
z_{1,2}\simeq 0$) and one heavy one $z_{3}\gg 1,$ resulting in a branching
ratio of 
\begin{eqnarray}
B\left( \mu e\gamma \right) _{W} &=&\frac{3\alpha }{8\pi }\left|
\sum_{i=1}^{3}\mathbb{U}_{\mu i}^{\ast }\mathbb{U}_{ei}F\left( \frac{%
m_{i}^{2}}{M_{W}^{2}}\right) \right| ^{2}  \notag \\
&=&\frac{3\alpha }{8\pi }\left| \mathbb{U}_{\mu 3}^{\ast }\mathbb{U}_{e3}%
\left[ F\left( \infty \right) -F\left( 0\right) \right] \right| ^{2}=\frac{%
3\alpha }{8\pi }\zeta ^{2}\theta _{L}^{4}
\end{eqnarray}
where $\zeta =%
{\frac12}%
\sin 2\omega ,$ as seen in Eq.(\ref{mixings})$.$ We have used the
orthogonality condition of the mixing matrix $\mathbb{U}$ when going to the
second line. Allowing for a large $\mu e$ mixing angle $\omega ,$ the
experimental limit\cite{muegammalimit} of $B\left( \mu e\gamma \right)
\simeq 10^{-11}$ requires a heavy-light angle $\theta _{L}\simeq
m_{1}/M_{1}=O\left( 10^{-2}\right) ,$ which is small, but still plausible as
we expect the Yukawa masses $m_{1}$ to be quite bit less than the Dirac
(bare) mass $M_{1}$. Thus the measured value begins to give meaningful
constraint on the model parameters. Our point is that the significant
restriction is on the mixing angle, rather than on the KK masses directly.
Note that if we had not taken into account the suppression due to the mixing
angles we would get an unacceptable large $B\left( \mu \rightarrow e\gamma
\right) .$ Also in this case the large mass comes from bare mass and not
from the large Yukawa coupling and we expect the decoupling to be valid\cite
{CL91}. Indeed $B\left( \mu \rightarrow e\gamma \right) $ vanishes in the
limit $M_{1}\rightarrow \infty ,$ after the behavior of the mixing angles is
included.

\section{$\protect\mu \rightarrow e\protect\gamma $ : heavy particle in the
scalar boson loop}

In the minimal SM with massive neutrinos, the leading $\mu e\gamma $
amplitude comes from the gauge boson loop as discussed in the previous
Section. However, for models having more scalars beyond the one Higgs
doublet, there could in principle be significant scalar boson loop
contribution as well. Even for the minimal SM, it is instructive to consider
the scalar boson case separately because the longitudinal gauge boson is
simply the (unphysical) Higgs scalar boson. Being proportional to the
fermion mass, such Yukawa coupling is the source of the decoupling violation
--- through the cancellation of the large mass in fermion propagator by the
large Yukawa couplings.

The Yukawa interactions of the scalar boson $\phi $ to a charged lepton 
\emph{l} and a massive neutrino $\nu _{i}$ can be parametrized by the chiral
couplings $y_{li}^{\left( \pm \right) }$: 
\begin{equation}
\mathcal{L}\left( \phi l\nu _{i}\right) =\bar{l}\left[ y_{li}^{\left(
+\right) }\left( 1+\gamma _{5}\right) +y_{li}^{\left( -\right) }\left(
1-\gamma _{5}\right) \right] \nu _{i}\phi _{{}}^{-}+h.c.
\end{equation}
We have performed a detailed calculation of the scalar boson loop amplitude, 
$\mu ^{-}\rightarrow \left( \nu _{i}\phi _{\gamma }^{-}\right) \rightarrow
e^{-}$, where the photon is emitted by the charged $\phi $ boson in the
intermediate state, and found that, for heavy neutrino intermediate state $%
\left( m_{i}\gg M_{\phi }\right) ,$ 
\begin{equation}
A_{+}^{\phi }\left( \nu _{\text{heavy}}\right) =\frac{1}{\pi m_{i}^{2}}%
\left( \frac{m_{\mu }}{3}y_{ei}^{\left( +\right) \ast }y_{\mu i}^{\left(
-\right) }+m_{i}y_{ei}^{\left( +\right) \ast }y_{\mu i}^{\left( +\right)
}\right) ,  \label{phi-heavy}
\end{equation}
and, for light neutrino intermediates state $\left( m_{i}\ll M_{\phi
}\right) $, 
\begin{equation}
A_{+}^{\phi }\left( \nu _{\text{light}}\right) =\frac{1}{\pi M_{\phi }^{2}}%
\left( \frac{m_{\mu }}{6}y_{ei}^{\left( +\right) \ast }y_{\mu i}^{\left(
-\right) }+m_{i}y_{ei}^{\left( +\right) \ast }y_{\mu i}^{\left( +\right)
}\right) .  \label{phi-light}
\end{equation}
The other chiral amplitudes $A_{-}^{\phi }$ have similar structures.

Let us first consider the SM case when the scalar is the would-be-Goldstone
boson, and becomes the longitudinal gauge boson after spontaneous symmetry
breaking. In the renormalizable $R_{\xi }$ gauge, there is a scalar particle
with a mass $M_{\phi }=\xi M_{W}.$ The SM fermions obtain their masses
through their couplings to the Higgs field, hence the Yukawa couplings are
proportional to the fermion masses. If the VEV is written in terms of $g$
and $M_{W}$, we have the explicit form of 
\begin{eqnarray}
y_{ei}^{\left( +\right) \ast } &=&\frac{gm_{i}}{2\sqrt{2}M_{W}}U_{ei}^{\ast
},\;\;\;\;\;\;\;\;\;y_{ei}^{\left( -\right) \ast }=\frac{-gm_{e}}{2\sqrt{2}%
M_{W}}U_{ei}^{\ast },  \notag \\
y_{\mu i}^{\left( +\right) } &=&\frac{gm_{\mu }}{2\sqrt{2}M_{W}}U_{\mu
i},\;\;\;\;\;\;\;\;\;y_{\mu i}^{\left( -\right) }=\frac{-gm_{i}}{2\sqrt{2}%
M_{W}}U_{\mu i},  \label{SMyukawa}
\end{eqnarray}
Substituting these relations into Eq.(\ref{phi-heavy}), we can check that
heavy neutrino $\left( \nu _{3}\right) $ contribution is given by 
\begin{equation}
A_{+}^{\phi }\left( \nu _{\text{heavy}}\right) =\frac{g^{2}m_{\mu }}{8\pi
M_{W}^{2}}\mathbb{U}_{\mu 3}^{\ast }\mathbb{U}_{e3}\left( \frac{2}{3}\right)
\end{equation}
in agreement with the result in Eq.(\ref{W-amplitude}) with $F\left( \infty
\right) =2/3.$ This shows that the heavy particle non-decoupling
contribution to the $\mu e\gamma $ amplitude comes entirely from the Higgs
boson loop\cite{CL91}.

For models with non-minimal Higgs structure, we have physical scalar
particles with couplings that do not have a simple fermion mass dependence
--- in fact they are as a rule highly model-dependent. The result of (\ref
{phi-heavy}) can then be translated, with $m_{\mu }/m_{i}\simeq 0,$ into the
branching ratio of 
\begin{equation}
B\left( \mu e\gamma \right) _{\phi \nu _{i}}=\frac{24\alpha }{\pi g^{4}}%
\left( \frac{M_{W}^{4}}{m_{\mu }^{2}m_{i}^{2}}\right) \left( y_{ei}^{\left(
+\right) \ast }y_{\mu i}^{\left( +\right) }\right) ^{2}.  \label{scalarBR}
\end{equation}
Clearly a naive assumption of $\left( y_{ei}^{\left( +\right) \ast }y_{\mu
i}^{\left( +\right) }\right) =O\left( 1\right) $ would lead to a
meaninglessly weak bound on the heavy neutrino mass\footnote{%
In this respect, we differ from the conclusion drawn in Ref. \cite{kitano}
where the scalar contribution to the branching ratio has been estimated to
have the mass dependence of $\left( M_{W}/m_{i}\right) ^{4}$, as compared to
our result of Eq.(\ref{scalarBR}). The author was also silent with regard to
the implication of the apparent decoupling violation by the W-loop
contribution, as stated in Eq.(\ref{W-amplitude}).} of $m_{i}>10^{7}TeV.\;$%
Since the generalized Yukawa couplings $y_{li}^{\left( \pm \right) }$ do
include small mixing angles, it seems more sensible to use the experimental
result\cite{muegammalimit} of $B\left( \mu e\gamma \right) \lesssim
1.2\times 10^{-11}$ to set a limit on the coupling and mixing angle
combination: 
\begin{equation}
\left( y_{ei}^{\left( +\right) \ast }y_{\mu i}^{\left( +\right) }\right)
\lesssim \frac{10^{-7}}{m_{i}\left( TeV\right) },
\end{equation}
where we have used the value of Fermi constant $G_{F}=\sqrt{2}%
g^{2}/8M_{W}^{2}\simeq 10^{-5}/M_{N}^{2}.$

\section{Discussion}

We have focused on the Randall-Sundrum version of the extra-dimensional
theory. However, our discussion is equally applicable to the original
version where the suppression of the bulk field effects (gravitons,
singlet-neutrinos, etc.) comes through the large volume of the extra
dimensional space. This is the case because the structure of the neutrino
mass matrix is very similar in both versions of the theory\cite{nuLargeVol}.

In this paper, we have concentrated on a single bulk neutrino. In principle,
there is a whole tower of Kaluza-Klein states. Many authors\cite{towerSUM}
have attempted to sum over the contribution by the whole tower. We have not
done so because we do not wish to confuse the issue of decoupling of a
single heavy particle with the separate problem of how the sum of this
infinite tower should behave. The individual heavy particle contribution is
controlled by the heavy-light mixing angle, which is mass-suppressed. If one
sums over an infinite number of such small terms, a ``non-decoupling''
result can be obtained. Clearly, this approach touches upon the difficult
issue of convergence of the KK sum, with implications related to the
possible presence of new physics at higher scales. Such problems are quite
different from the matter of single particle decoupling, which is the focus
of this paper.

One of us (T.P.C.) would like to thank Gary Shiu for helpful discussion.
L.F.L. acknowledges the support from U.S. Department of Energy (Grant No.
DE-FG02-91ER40682).

\end{document}